# The impact of the microvascular resistance on the measures of stenosis severity

Tam Atkins [1*], Navid Freidoonimehr [1], John Beltrame [2,3,4], Christopher Zeitz [2,3,4], Maziar Arjomandi [1]

[1] School of Electrical and Mechanical Engineering, The University of Adelaide, Adelaide, SA 5000, Australia

[2] School of Medicine, Faculty of Health Sciences, The University of Adelaide, Adelaide, SA 5000, Australia

[3] Central Adelaide Local Health Network, Adelaide, SA 5000, Australia

[4] Basil Hetzel Institute for Translational Health Research, Adelaide, SA 5011, Australia

[*] Corresponding Author: a1706581@adelaide.edu.au

# Abstract

The relationship between measures of stenosis and microvascular resistance is of importance due to medical decisions being based on these values. This research investigates the impact of varying microvascular resistance on fractional flow reserve (FFR) and hyperaemic stenosis resistance (hSR). Microvascular resistance is classified using hyperaemic microvascular resistance (hMR). Additionally, hMR using the upstream pressure value ($hMR_{Pa}$) has also been calculated and is compared to hMR measured conventionally. Tests were conducted at three different degrees of stenosis (quantified by percent area) in a coronary flow circuit with varying downstream resistance to simulate the microvasculature. Pressure and flow values are recorded across the stenosed section, allowing for calculation of the diagnostic indexes. Results indicate that for a constant degree of stenosis, FFR would increase with increasing microvascular resistance while hSR would remain almost constant. $hMR_{Pa}$ was found to approach hMR as the stenosis severity decreased, and the pressure gradient decreased. In the results shown here, with sufficiently high downstream resistance, an 84% stenosis could produce an FFR value over 0.8. This result suggests that there is the potential for misdiagnosis of the severity of stenosis when combined with elevated microvascular resistance. Consequently, decisions on the clinical significance of a stenosis, classified by FFR, need to consider the effect of the microvascular resistance.

# Introduction

Coronary arteries are the vessels that carry oxygenated blood to the heart muscle. Oxygenated blood travels through the coronary arteries and through branching vessels of increasingly smaller diameter until it reaches the microvasculature. The microvasculature refers to the extremely small vessels that perfuse oxygenated blood to the heart that are too small to be seen using conventional imaging techniques (Ong et al. 2018; Vancheri et al. 2020).

Broadly, there are two coronary causes of limited blood flow to the heart muscle. One is obstruction of the coronary artery and the other is dysfunction of the microvasculature. Coronary artery disease is typically caused by a build-up of atherosclerotic deposits on the artery walls (Mendis et al. 2011). These deposits increase in size until they obstruct coronary blood flow and produce symptoms such as angina and shortness of breath (Cassar et al. 2009). Coronary microvascular dysfunction (MVD) occurs when the microvasculature itself is responsible for the lack of blood to the heart muscle, rather than blockages of the coronary arteries (Vancheri et al. 2020). This covers a number of disorders and can be separated into types based on combinations with other diseases (for example, Camici and Crea (2007) defined a disorder type based on the combination of MVD with coronary artery stenosis). Generally, microvascular dysfunction is not caused by blockages within the microvasculature (Vancheri et al. 2020) but by inflammation related factors or conditions that limit the ability of the vessels to dilate to deliver more blood as required.

Patients exhibiting symptoms of coronary artery stenosis typically undergo an angiogram to evaluate the clinical status of the arteries. If the angiogram detects a narrowing of the arteries greater than 75% (although this percentage varies in subsequent literature, as noted in Vancheri et al. (2020)), then the stenosis is considered significant. Work by White et al. (1984) suggested that angiographic classification of stenosis was often inaccurate, and, for stenosis over 60%, their severity was frequently underestimated. More recently, other methods of classifying the severity of a stenosis have been developed.

Fractional flow reserve (FFR) is commonly used to ascertain whether a stenosis is functionally obstructive or not. This is measured as the ratio of the mean pressure measured downstream of the stenosis to the mean pressure measured upstream (Pijls et al. 1996). Generally, the cut off value for intervention is an FFR value of 0.75 or 0.8 (Mohdnazri et al. 2016), below which patients are considered for revascularisation therapies (stenting or bypass surgery). Another measure of the stenosis severity is the hyperaemic stenosis resistance (hSR), defined as the ratio between the pressure gradient across a stenosis and the hyperaemic distal blood velocity (Meuwissen et al. 2002). An unobstructed artery will have a hSR value of approximately 0, as the pressure gradient will be almost zero. Research shows a cut off value for hSR of 0.8 mmHg/cm/s (Meuwissen et al. 2002).

In the absence of a significant narrowing on angiography, the cause of the patient's symptoms may be due to microvascular dysfunction. The function of the microvasculature can be assessed using a pressure-temperature wire or combined Doppler-pressure wire. In the past, patients who displayed symptoms of myocardial ischaemia, such as angina, but had normal angiograms were discounted as having a cardiac problem (Beltrame et al. 2002; Vancheri et al. 2020). However, research now suggests that these patients can be experiencing microvascular dysfunction and patient studies have shown that measures of microvascular resistance (MVR) can be predictors of clinical outcomes in these patients. Several indexes can be used to quantify the MVR invasively. The two more common measures of MVR are the index of microvascular resistance (IMR) and hyperaemic microvascular resistance (hMR). IMR is measured using a pressure/temperature sensing wire (De Bruyne et al. 2001), allowing calculation of the travel time of room temperature saline. The distal pressure measurement, divided by the inverse of the mean travel time, provides the value of IMR (Fearon et al. 2003). hMR is found using a combined pressure/Doppler wire, capable of measuring both pressure and velocity simultaneously. hMR is the ratio of downstream pressure to the velocity measurement. Both indexes are measured at hyperaemia and hence the minimum values of resistance are being recorded. A comparison between hMR and IMR was conducted by Demir et al. (2022), who suggested that there was a weak correlation between the two indexes.

The measurement of FFR can be affected by a number of parameters, these parameters can have varying effects on the accuracy of simulations (Sankaran et al. 2016). Fearon et al. (2004) showed that if collateral flow was accounted for then there was limited relation between the measure of FFR and IMR. In patients, van de Hoef et al. (2014) found that an elevated hMR value would cause an increase in the FFR for a constant hSR. Eftekhari et al. (2022) showed that the relationship between FFR, hSR and hMR could be related as shown in Equation 1.

$$FFR = \frac{hMR}{hMR + hSR} \quad (1)$$

In-vitro test setups for simulating the coronary flow do not always include a separate component to simulate the MVR. Freidoonimehr et al. (2020b) investigate flow through stenosed artery models without including a component to mimic the MVR. However, the work of Kolli et al. (2022) used a needle valve to provide variable MVR.

This research investigates the effect of varying downstream resistance due to changes in the microvasculature on measures of stenosis severity, using a specialised experimental apparatus allowing for the control and controlled variation of specific parameters, in a way not possible in existing in-vivo studies. For controlled stenosis models, downstream MVR is increased and the effect on FFR and hSR is recorded. This data is, for the first time, used to systematically compare the effectiveness of FFR and hSR for predicting stenosis severity in the presence of elevated MVR. Finally, the data is validated by comparison with relationships generated in recent literature, and then used to extend recent research into the location of the pressure wire during cardiac catheterisation to include the effect of an upstream stenosis.

# Method

## Test apparatus

The test apparatus consists of a flow circuit with the flow behaviour controlled by a programmable centrifugal pump (RS Components) capable of delivering a steady or oscillatory flow up to 500 ml/min. The circuit is formed from 6 mm inner diameter acrylic pipe, with a long (approximately 2 m) straight section before the test section to ensure flow is fully developed. Upstream of the test section an electromagnetic flow meter (Krohne BATCHFLUX 5500 C) is mounted to record the flow rate entering the test section. Pressure sensors (Druck UNIK 5000) are mounted either side of the test section. At the end of the flow circuit and just before the pump, a reservoir holds the fluid. The sensor output and pump input are all managed through LabVIEW software (National Instruments). The setup is shown in Figure 1. From Grist et al. (1997), the average hyperaemic flow rate has been taken as 234 ml/min. Due to the internal diameter used here, matching the Reynolds number gives a hyperaemic flow rate of 440 ml/min. The tested flow range was approximately 200-460 ml/min.

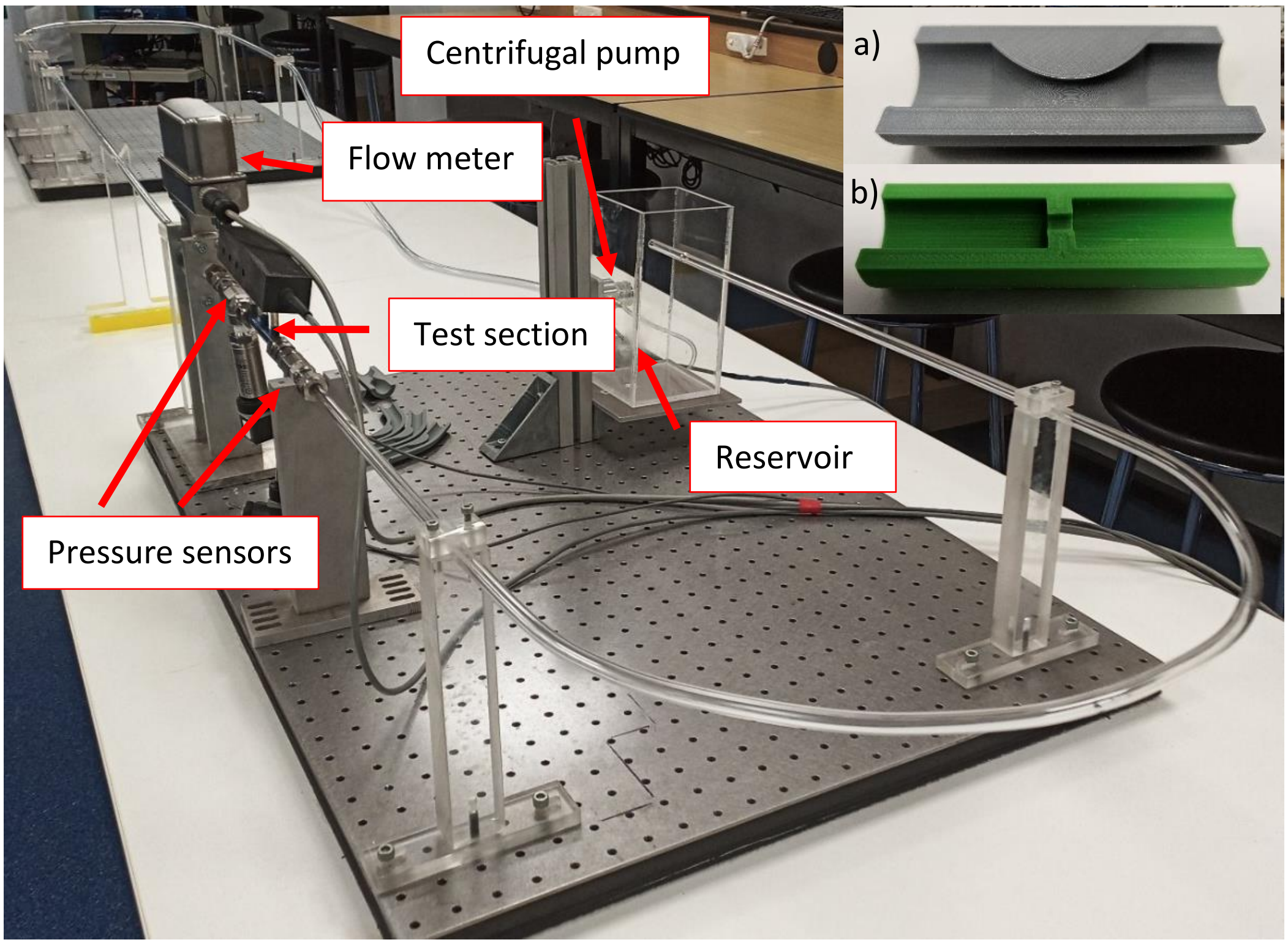

Figure 1: Main image: Test apparatus with labels showing key features. Inset images: a) cross section of the stenosis models, b) cross section of the orifices used to simulate microvascular resistance.

## Models of stenosis and flow restrictors

Manufacture of the stenosis models used here has been described previously in Freidoonimehr et al. (2020b) and is briefly summarised here. The stenosis models are made from 3D printed acrylonitrile butadiene styrene (ABS) and have been treated in an acetone vapour chamber to smoothen the surface. For the purposes of this research, eccentric stenoses were used (profile shown in Figure 1a), in three different area percent stenosis, 64, 75 and 84%.

MVR in this setup is modelled as a flow restriction downstream of the stenosed section. These flow restrictions are 3D printed, square edged orifices (profile shown in Figure 1b) with varying bore size. The sizes range from 1.5 mm to 4 mm bore, shown in Table 1. The sizes were selected by a trial and error method to produce a range of hMR values covered in literature (Eftekhari et al. 2022; Feenstra et al. 2023). Using ABS, the orifices were 3D printed horizontally in a Zortrax M200 printer. Once printed, the ends of the orifices were filed smooth on the outside and precision drill bits were used to ensure the orifice bore was the correct diameter. Finally, the orifices were suspended and syringed through with acetone to smoothen the internal surface. Once completed, the orifices were mounted approximately 30 cm (50D) downstream of the stenosis. This distance ensures that the flow is laminar and fully developed after the stenosis and prior to the orifice and is substantially beyond the 10D likely secondary stenosis location (Freidoonimehr et al. 2020a). Modelling the MVR as a single restriction is justified by existent literature on branching and stenoses (Daniels et al. 2012) and is discussed in the Appendix. Unlike the microvasculature, this configuration will produce a constant resistance, which is reasonable given the constant flow rate investigated here. The downstream pressure sensor is mounted between the stenosis and the orifice, to mimic clinical pressure measurement.

## Experimental test configurations

Each test case corresponds to a different combination of stenosis severity and downstream orifice size. Test cases are outlined in Table 1. Five flow rates in the given range were tested. Pressure sensor orientation and spacing was controlled during testing to minimise random error. Each data point is an average of five tests. Once collected, data was processed in MATLAB 2021b. A detailed analysis of sensor error and standard deviation is covered in the Appendix.

Table 1: Test cases used to collect the data presented in this work.

| Stenosis model | Downstream orifice | Symbol |
|---|---|---|
| No stenosis | 1.5mm | ◊ |
| | 1.8 mm | △ |
| | 2 mm | X |
| | 3 mm | O |
| | 4 mm | □ |
| 64% | No orifice | * |
| | 1.5 mm | ◊ |
| | 1.8 mm | △ |
| | 2 mm | X |
| | 3 mm | O |
| | 4 mm | □ |
| 75% | No orifice | * |
| | 1.5 mm | ◊ |
| | 1.8 mm | △ |
| | 2 mm | X |
| | 3 mm | O |
| | 4 mm | □ |
| 84% | No orifice | * |
| | 1.5 mm | ◊ |
| | 1.8 mm | △ |
| | 2 mm | X |
| | 3 mm | O |
| | 4 mm | □ |

In each test case (except the test cases using the orifices alone) the FFR, hSR, hMR and $hMR_{Pa}$ were calculated, using equations 2 to 5. *Pa* is the aortic pressure, approximated as the pressure upstream of the stenosis. *Pd* is the pressure downstream of the stenosis. *V* is the velocity of the fluid in the test section, found by dividing the flow rate by the cross-sectional area of the unobstructed pipe.

$$FFR = \frac{P_d}{P_a} \qquad (2)$$

$$hSR = \frac{(P_a - P_d)_{mean}}{V_{mean}} \qquad (3)$$

$$hMR = \frac{P_d}{V_{mean}} \qquad (4)$$

$$hMR_{Pa} = \frac{P_a}{V_{mean}} \qquad (5)$$

## Results

For each orifice and stenosis case, standalone tests were completed and the pressure drop at each flow rate was recorded. As shown in Figure 2a, the larger the area of blockage, the larger the pressure drop. For example, at a flow rate of approximately 200 ml/min, the 64% stenosis is producing a pressure drop of around 0.9 mmHg, the 75% stenosis a pressure drop of around 1.2 mmHg and the 84% stenosis a pressure drop of around 3 mmHg. As the severity of the blockage increases, the relationship between pressure drop and flow rate increases in slope.

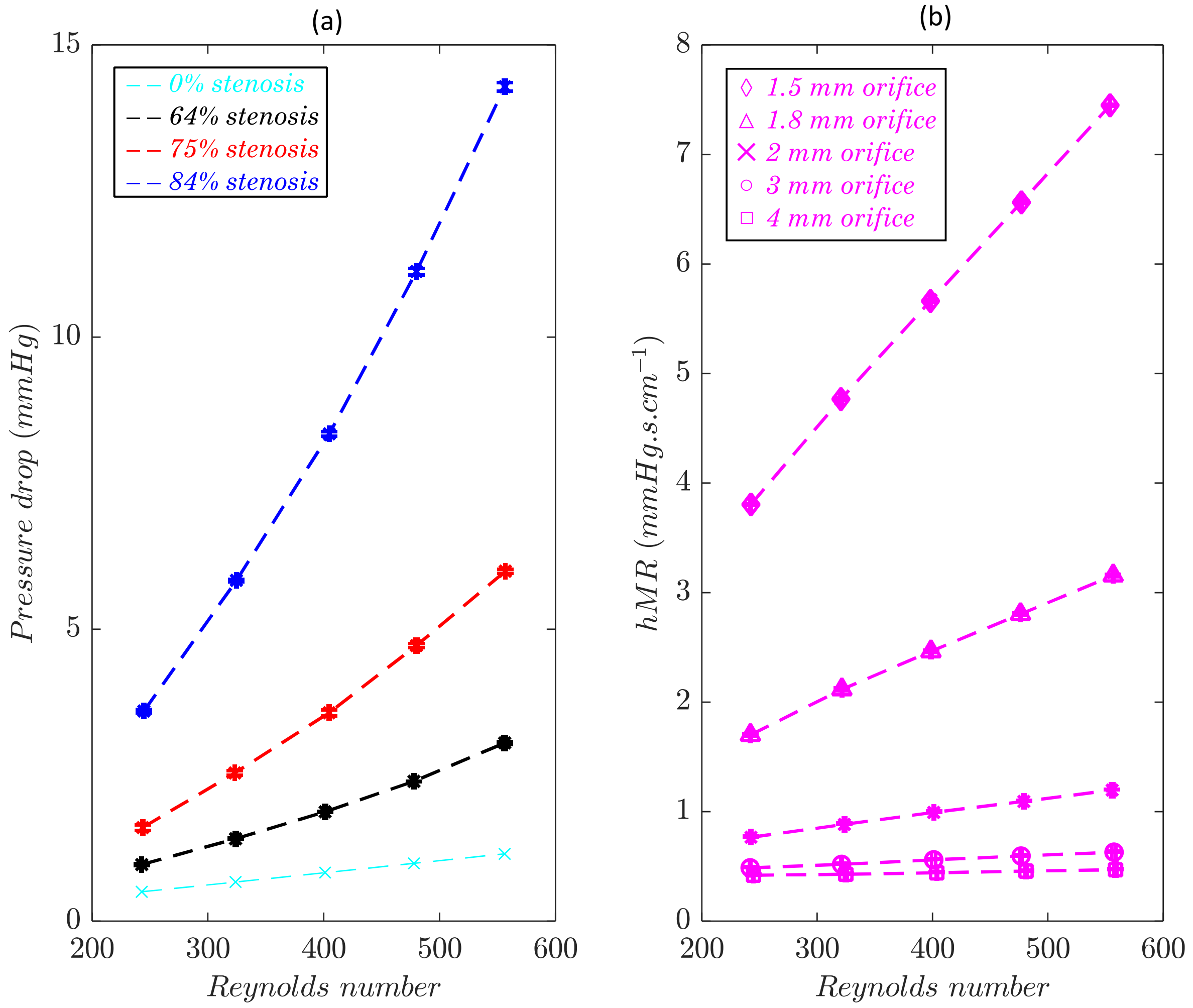

Figure 2: (a) Pressure drop at different Reynolds number for the three different stenosis cases (with no downstream orifice) used here. The light blue line is 0% stenosis (i.e. a pipe), estimated using the Darcy Weisbach formula, the black line represents the 64% stenosis, the red line the 75% stenosis and the dark blue line the 84% stenosis. (b) hMR at different Reynolds number for the five different orifices (with no stenosis upstream) used to create microvascular resistance here. ◊denotes an orifice of 1.5 mm diameter, △ denotes an orifice of 1.8 mm diameter, X denotes an orifice of 2 mm diameter, O denotes an orifice of 3 mm diameter, ☐ denotes an orifice of 4 mm diameter. Error bars are shown for both the pressure drop and the flow rate

The hMR due to the orifices is presented as an average of all cases against Reynolds number in Figure 2b. hMR increases with decreasing bore diameter and the rate of hMR increase grows as the diameter decreases. This is demonstrated by comparing the 3 and 4 mm cases with each other; and the 1.5 and 1.8 mm cases with each other. For example, at a Reynolds number of approximately 400, the difference in hMR between the 3 and 4 mm orifices is approximately 0.1 mmHg.s.cm$^{-1}$ and the difference in hMR between the 1.5 and 1.8 mm cases is approximately 3 mmHg.s.cm$^{-1}$.

## Combined stenosis and microvascular resistance

Each stenosis model was tested in combination with each orifice downstream. From this, the pressure drop across the stenosed section was calculated in each case. For the 75% stenosis case, the pressure drop and flow rate data are shown in the Appendix. The hMR value does not change significantly with changing stenosis upstream, as shown in Figure 3. This suggests that the orifices are able to produce a constant microvascular condition.

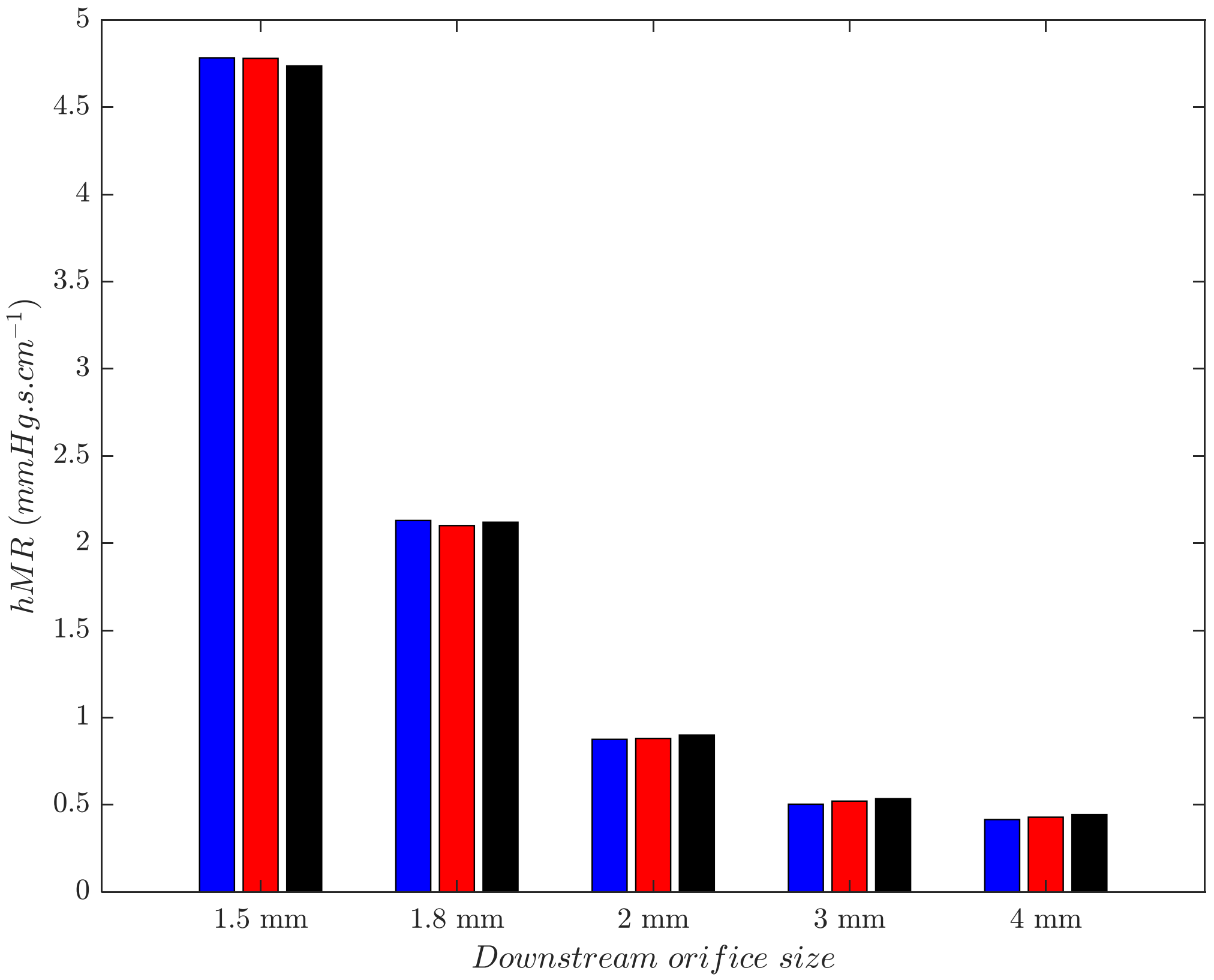

Figure 3: Each orifice is capable of producing an almost constant hMR value, despite the changes in stenosis upstream. For each group of bars, the black denotes the 64% stenosis the, the red the 75%

stenosis and the blue the 84% stenosis. The groups are based on the downstream orifice size, as noted on the x-axis. This figure is produced holding the flow rate constant at approximately 280 ml/min.

## Effect of changing downstream resistance on FFR

Changing the resistance downstream caused a change in the FFR value for each stenosis. Figure 4 shows the FFR calculated across the stenosis in each test configuration, with each subplot showing a different stenosis. For each case, the increasing downstream pressure due to the decreasing orifice bore size causes a shift in the FFR line upwards. Based on the definition of FFR, at a higher flow rate and higher MVR, both Pa and Pd are increased by a constant pressure, which results in an increase in the FFR value.

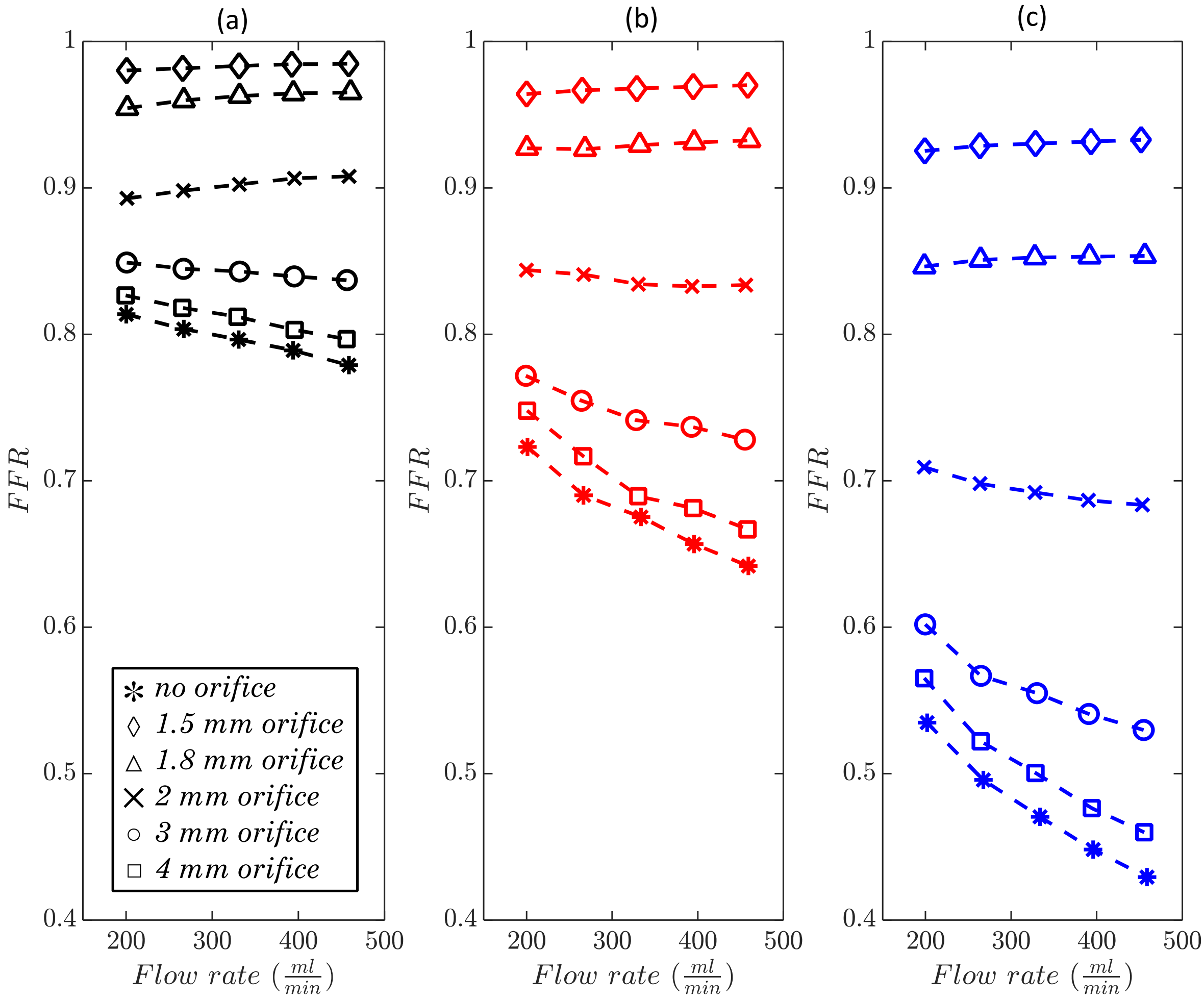

Figure 4: FFR versus flow rate for each stenosis with varying downstream orifice size. The black lines represent 64% stenosis (a), the red 75% stenosis (b) and the blue 84% stenosis (c). * denotes with no orifice downstream, ◊denotes an orifice of 1.5 mm diameter, △ denotes an orifice of 1.8 mm diameter, X denotes an orifice of 2 mm diameter, O denotes an orifice of 3 mm diameter, ☐ denotes an orifice of 4 mm diameter.

## Effect of changing downstream resistance on hSR

Changing the downstream resistance does not significantly alter the hSR across a stenosis. Figure 5 shows the hSR calculated for each case tested here, each subplot corresponds to a different stenosis severity. hSR is altered by flow rate but is largely unaltered by changing downstream resistance. The slope of the linear relationship between hSR and flow rate varies with varying degree of upstream stenosis.

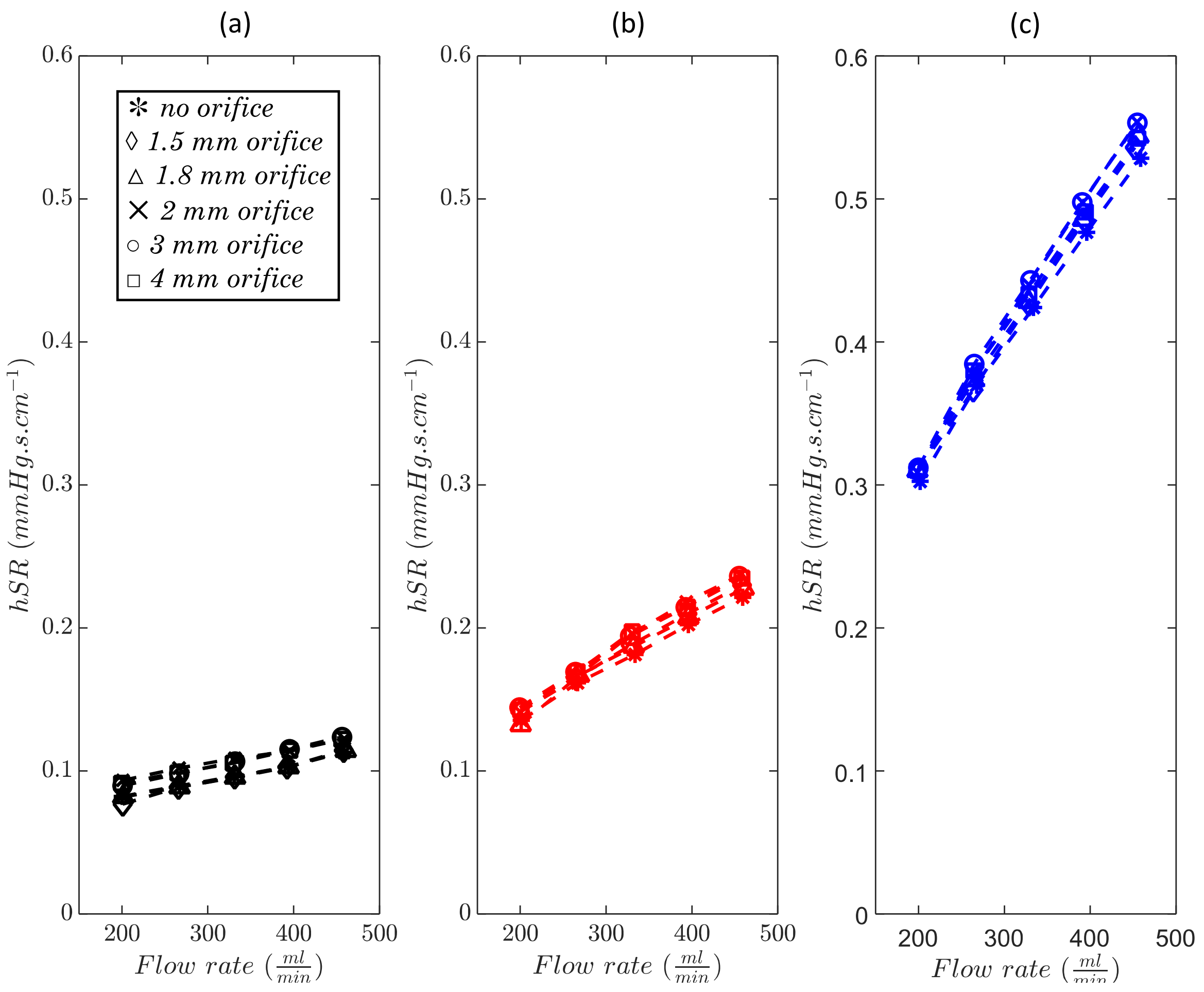

Figure 5: hSR vs flow rate for each stenosis with varying downstream orifice size. The black lines represent 64% stenosis (a), the red 75% stenosis (b) and the blue 84% stenosis (c). * denotes with no orifice downstream, ◊ denotes an orifice of 1.5 mm diameter, △ denotes an orifice of 1.8 mm diameter, X denotes an orifice of 2 mm diameter, O denotes an orifice of 3 mm diameter, ☐ denotes an orifice of 4 mm diameter.

## Location of measuring hMR

In the absence of a stenosis, hMR can be calculated using *Pa* rather than *Pd* (Feenstra et al. 2023). $hMR_{Pa}$, calculated using the upstream pressure as an approximation for *Pa*, was compared with the hMR calculated using *Pd* (Figure 6). As the severity of the stenosis decreases, the relationship between $hMR_{Pa}$ and hMR approaches equality and thus the results of Feenstra et al. (2023). Increasing pressure downstream of the stenosis due to decreasing orifice bore size causes the line formed by the two measures of hMR to approach the slope of 45°.

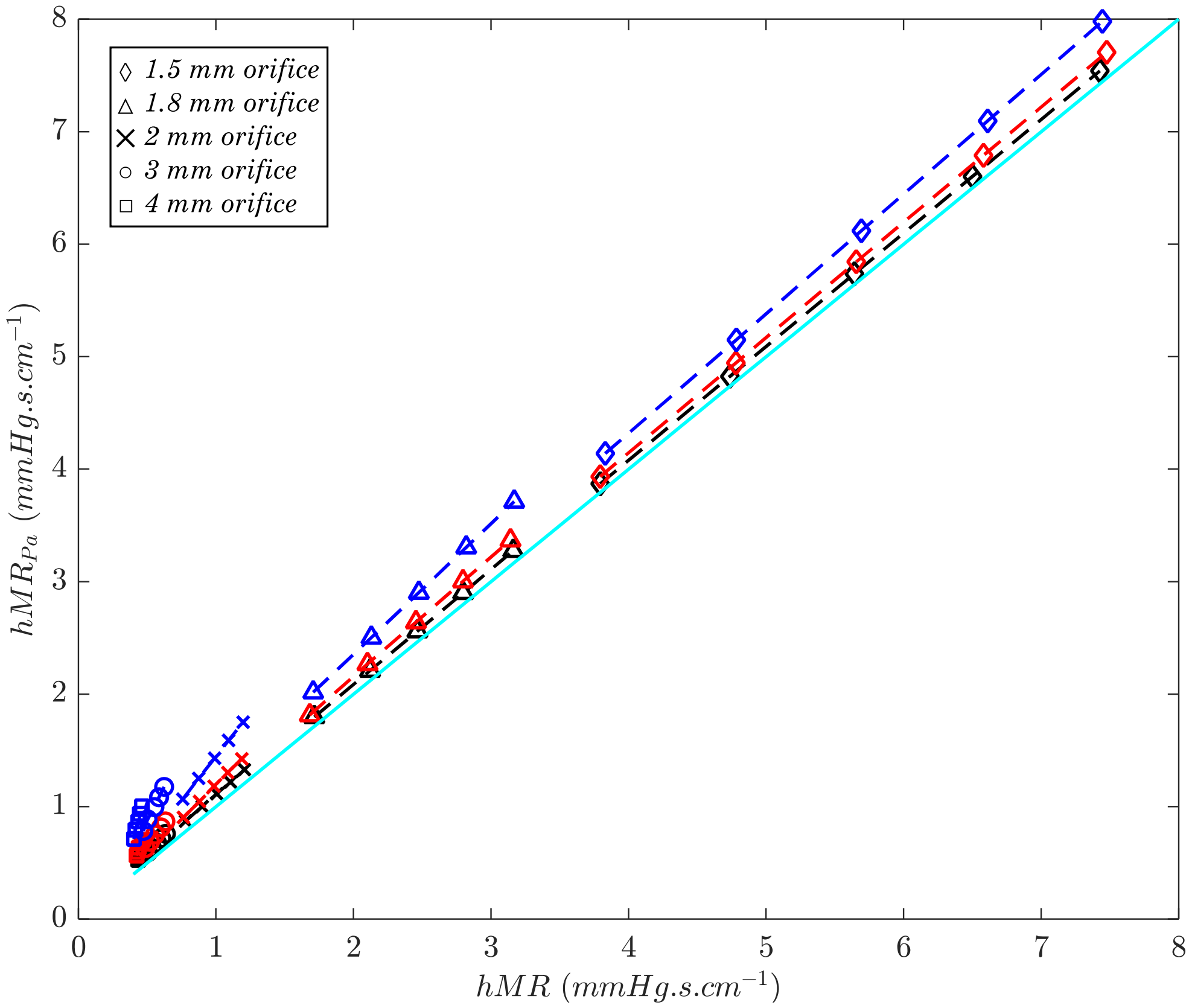

Figure 6: Comparison between hMR calculated using upstream pressure ($hMR_{Pa}$) and hMR calculated using downstream pressure. The black dashed line represents 64% stenosis, the red 75% stenosis, the dark blue 84% stenosis and the solid light blue the line of y = x. * denotes with no orifice downstream,

◊ denotes an orifice of 1.5 mm diameter, △ denotes an orifice of 1.8 mm diameter, X denotes an orifice of 2 mm diameter, O denotes an orifice of 3 mm diameter, ☐ denotes an orifice of 4 mm diameter.

# Discussion

These tests have shown that measuring the functional significance of stenoses can be affected by variations in the microvasculature. The results suggest that FFR is directly altered by increasing MVR. However, hSR does not appear to be affected directly by the increase of MVR downstream. Measuring hMR using upstream pressure can be affected by the presence of a stenosis.

## Relationship between FFR and hMR

Figure 7 shows the relationship between the measures of FFR and hMR at three fixed flow rates. The flow rates are fixed at approximately 280 ml/min, 340 ml/min and 460 ml/min. As in Figure 4, these plots are largely independent of flow rate. Figure 7 shows that despite a constant area stenosis, the FFR reading across the stenosis increases with increasing hMR. Hence, FFR may not be an indicator of occlusion amount in the case of sufficiently elevated hMR. Normal values for hMR vary in literature but assuming a healthy cut off of 2.5 mmHg/cm/s (Feenstra et al. 2023) and a cut off value of FFR of 0.75 - 0.8 (Mohdnazri et al. 2016), it can be seen that all the stenoses presented here would present in the range of acceptable FFR when combined with elevated hMR, despite the high percent area stenosis.

The shaded red regions in Figure 7 show the range of FFR values that clinical decisions are made for. In sub-figure (b), for the 84% stenosis, at an hMR value below 2, the FFR value is below 0.75 and would indicate a need for clinical intervention. For a short window around an hMR value of 2, the patient would be in the clinical grey area and for an hMR of 2.5 and above, the FFR value would indicate insignificant stenosis, despite the 84% stenosed area. This result suggests that in the presence of a significantly elevated hMR, a correction factor may need to be applied to an FFR reading to accurately quantify the severity of the stenosis.

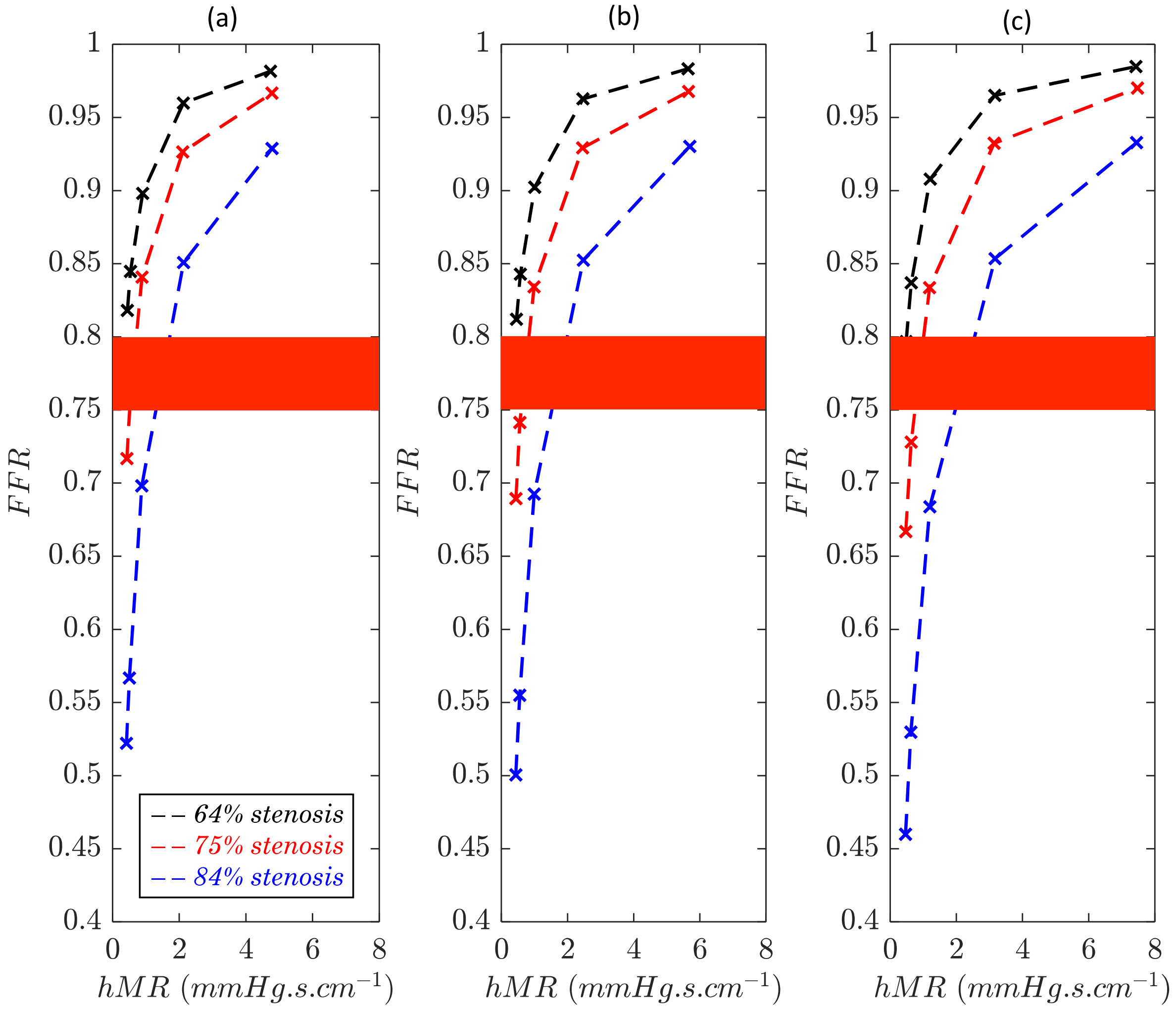

Figure 7: FFR vs hMR at three fixed flow rates. The black lines represent 64% stenosis, the red 75% stenosis and the blue 84% stenosis. a) is at approximately 280 ml/min, b) at approximately 340 ml/min and c) at approximately 460 ml/min flow rate. The shaded red regions show the range of cut off values (Mohdnazri et al. 2016).

## Relationship between hSR and hMR

Figure 8 shows hSR for each stenosis case with changing hMR value due to increasing downstream resistance. This figure shows that in each case, despite large changes in the value of hMR, the value of hSR remains relatively constant, and distinct for each stenosis severity. This is supported by the work of Eftekhari et al. (2022), who suggested that the measurement of hSR was more independent of the behaviour of the microvasculature than the measurement of FFR when quantifying the severity of a stenosis.

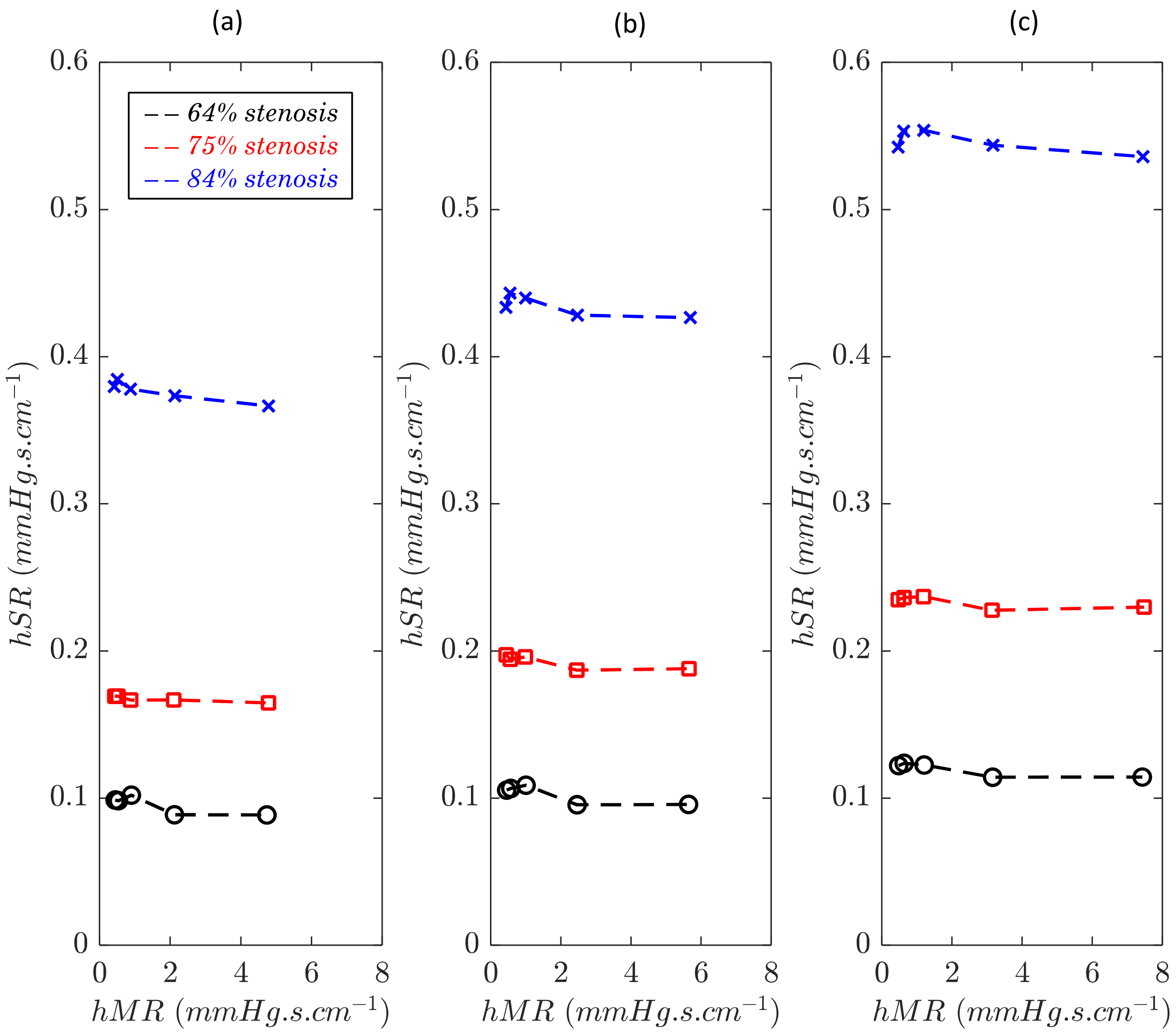

Figure 8: hSR vs hMR downstream at three different flowrates, for each stenosis type. The black line represents 64% stenosis, the red 75% stenosis and the blue 84% stenosis. a) is at approximately 280 ml/min, b) at approximately 340 ml/min and c) at approximately 460 ml/min flow rate.

Based on the definition of hSR (Equation 3), the value of the hSR is dependent on the velocity of the flow, however there are weak changes with varying hMR. The definitions for both can be rearranged for velocity and then equated, producing the relationship shown in Equation 6. Equation 6 shows a weak relationship that is dependent on a large change in the pressure ratio to produce a significant effect on hSR for a constant hMR value. This suggests that in the presence of an elevated hMR, hSR may be a better true indication of the functional severity of a stenosis than FFR.

$$hSR = hMR\left(\frac{P_a}{P_d} - 1\right) \qquad 6$$

## Importance of measuring location of hMR in presence of stenosis

$P_a$ is always higher than $P_d$, as a result of this, $hMR_{Pa}$ will always be larger than the corresponding value of hMR. However, in the absence of a stenosis, the difference between $P_a$ and $P_d$ will be small and due to viscous pressure losses alone. Hence $hMR_{pa}$ will be almost equal to hMR. When a stenosis is present, the value of $P_d$ becomes significantly lower than that of $P_a$, leading to a more significant reduction in the hMR value when compare with $hMR_{Pa}$. In Figure 6 this behaviour can be seen in all cases of varying downstream resistance and the 84% stenosis tests have $hMR_{Pa}$ values that are furthest from the values of hMR. For cases with less downstream resistance due to a larger orifice diameter, the slope of the relationship between hMR and $hMR_{Pa}$ is noticeably steeper than cases with higher downstream resistance. This is due to the dominant overall pressure increase in the presence of a severe downstream resistance.

The mathematical relationship shown in Equation 1 has been plotted in Figure 9a and demonstrates that the data acquired here displays the same trend that was displayed in human studies by Eftekhari et al. (2022). Figure 9a shows a clear relationship between these quantities, with the data points lying along the line y = x. Furthermore, a graph with $hMR_{Pa}$ substituted for hMR was plotted (Figure 9b) and shows that as downstream resistance increases, the results approach that of Figure 9a. This is due to the overall increase in pressure causing a proportionally smaller pressure drop across the stenosis (that is, *Pd* approaches *Pa*). Hence, the FFR value increases and $hMR_{Pa}$ approaches hMR. Figure 9b demonstrates the importance of consistent location of measurement in the presence of a stenosis.

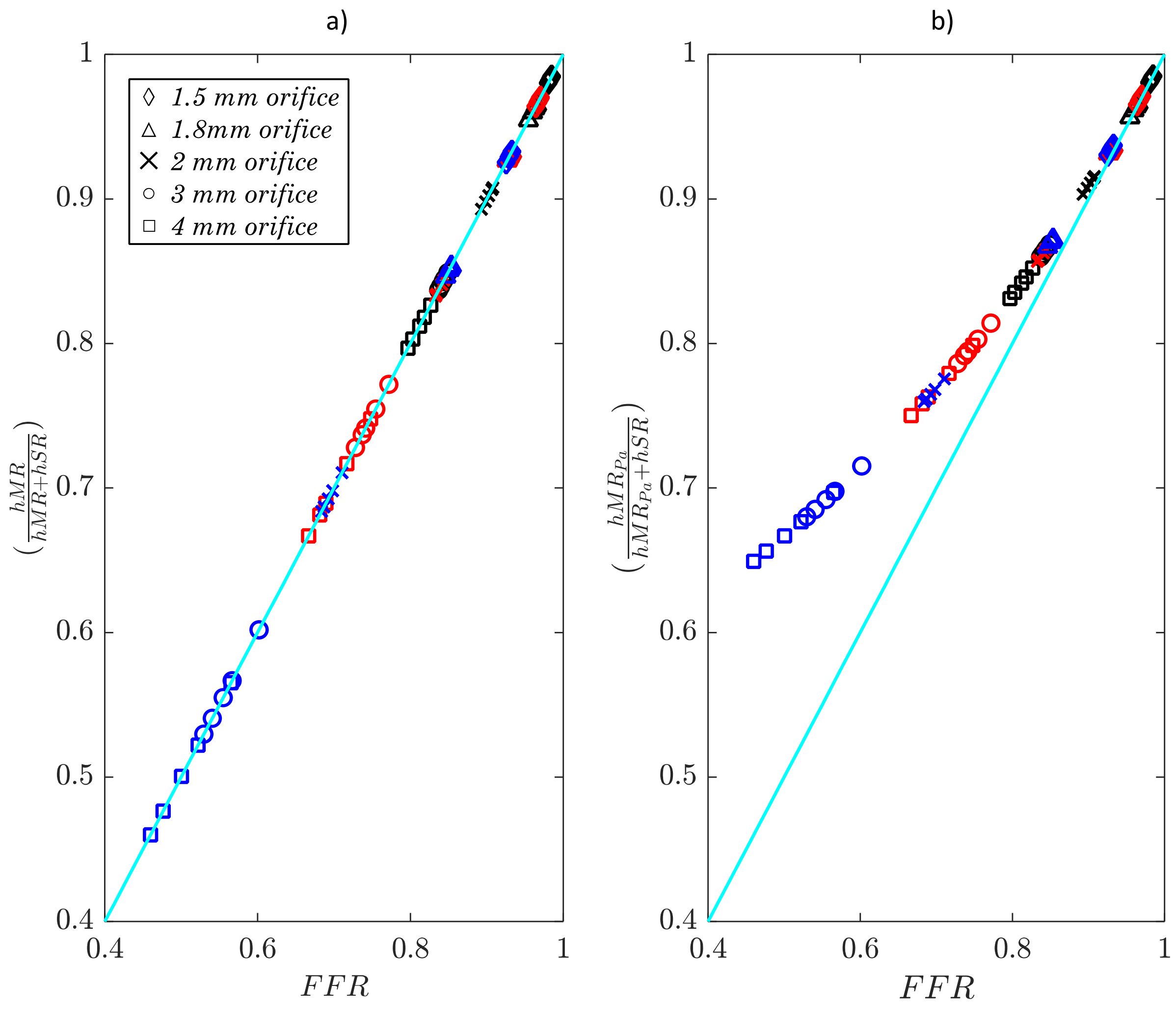

Figure 9: Plots of hMR/(hMR+hSR) vs FFR for each test case, a) and $hMR_{Pa}/(hMR_{Pa}+hSR)$ vs FFR for each test case, b). The black represents 64% stenosis, the red 75% stenosis, the dark blue 84% stenosis and the light blue solid line the line of y = x. * denotes with no orifice downstream, ◊ denotes an orifice of 1.5 mm diameter, △ denotes an orifice of 1.8 mm diameter, X denotes an orifice of 2 mm diameter, O denotes an orifice of 3 mm diameter, ☐ denotes an orifice of 4 mm diameter.

## Conclusions

This research was undertaken to develop an understanding of the effect of MVR on two indexes used to classify stenosis severity in a medical setting. FFR is typically used to assess the severity of a stenosis without consideration for the microvascular function. Previous research suggested that there was a relationship between the value of the MVR and the value of FFR. Research also suggested that classifying stenosis using hSR may be more independent of the MVR. The experiments completed here investigated these statements using an in-vitro set up that allowed parameters to be controlled and methodically varied. Additionally, the effect of the location of the measurement of hMR, in relation to an upstream stenosis, on the value of hMR was investigated as this was not found in existing literature.

FFR, hSR, hMR and $hMR_{Pa}$ were calculated from a combination of 3 upstream stenosis and 5 downstream microvascular resistors. The results showed that values of FFR can be almost doubled by increasing hMR. For example, at a hypaerimc flow rate, the FFR value can change from 0.46 to 0.92 for a change in hMR of 0.5 to 7.5 mmHg/s. This results in severe obstructions (when classified by percent area) producing an FFR reading indicative of a smaller obstruction i.e. over the cut-off value of 0.8. hSR is almost unchanged by altering the downstream resistance with a maximum change of approximately 7% across the range of hMR tested here. Measuring hMR upstream of a stenosis causes a shift in hMR reading, which increases in the case of severe stenosis due to the pressure drop across the stenosis. The mathematical relationship between the FFR, hMR and hSR that was derived in previous research was shown to hold true for the data collected.

The results presented in this work suggest that measuring the severity of a stenosis using FFR alone may lead to incorrect classification in the presence of elevated MVR. This may suggest that in patients with known elevated MVR, an alternative measure of stenosis (such as hSR) may be more effective than FFR. Further research into a correction factor that could be applied to FFR in the case of a known hMR may be warranted.

# Acknowledgements

Australian Government Research Training Program Scholarship.

# Conflict of interest statement

Nothing to disclose.

# Appendix

## Resistance Model

In the model considered here, the microvasculature is treated as an unobstructed branching pipe network, which is modelled as a number of resistors in parallel. Figure A1 shows the progression of simplifying the microvasculature to a number of resistors in parallel. These resistors can be combined into a single resistance, $R_m$, using the laws for parallel resistors. A single resistor upstream is used to model the stenosis. The single combined microvascular resistance from Figure A1, $R_m$, is applied through one orifice.

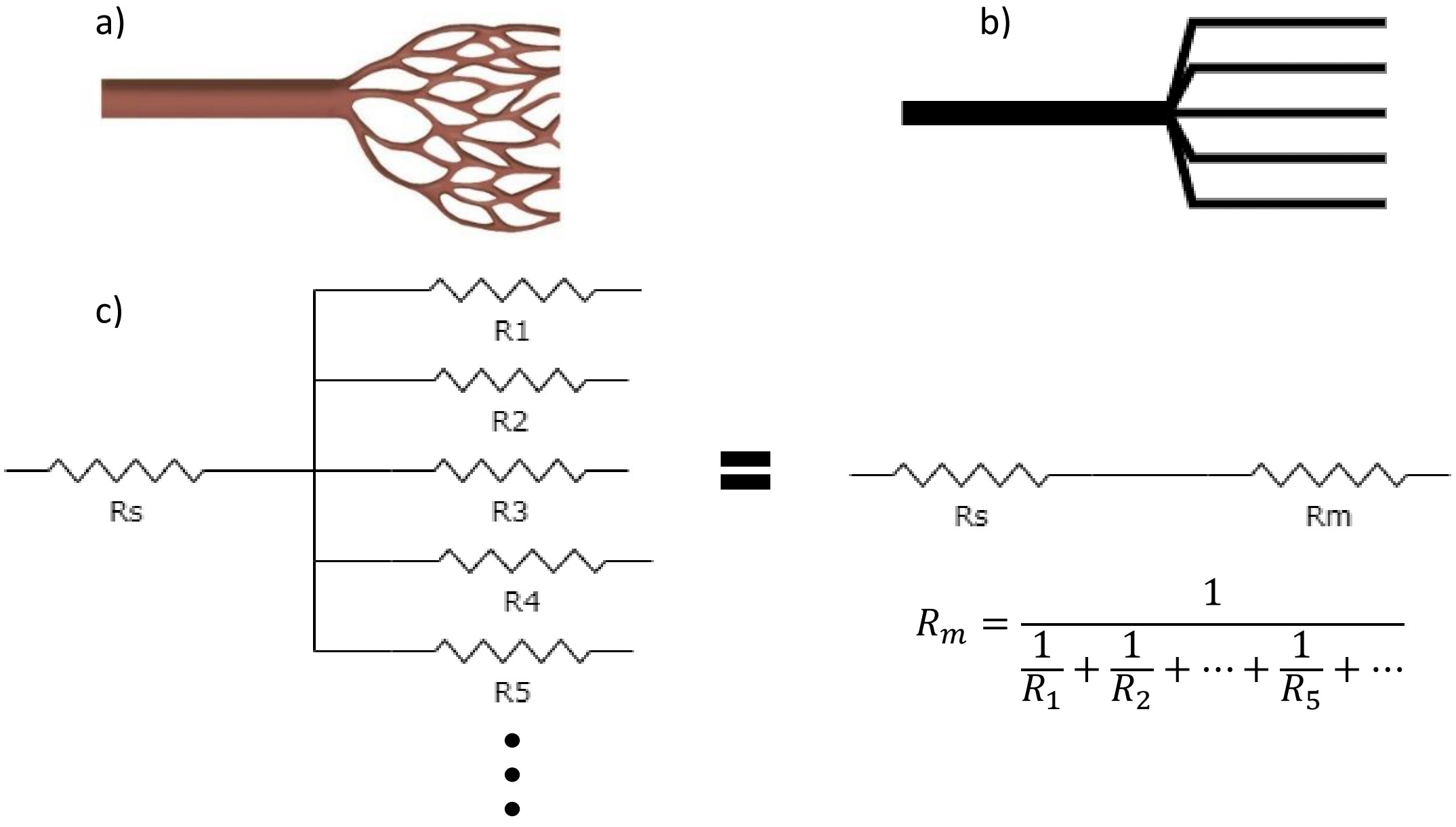

Figure A1: Showing the progression from the human model a), to the simplified model b) to the resistor approximation c). Resistors $R_1$ – $R_5$ represent microvascular vessels, these are summed using the equation shown to develop the overall resistance due to the microvasculature, $R_m$, and resistor $R_s$ represents the coronary artery stenosis.

## Pressure drop across stenosis

Varying the orifice size downstream of the stenosed section had only a small effect on the pressure drop across the stenosed section (Figure A2). For example, at a flow rate of approximately 260 ml/min, the pressure drop ranges from 2.5 mmHg to 2.6 mmHg. As the flow rate increased, the pressure drop values become more distinct for each case. This is shown at the highest flow rate (approximately 460 ml/min), where the pressure drop varies from 6 mmHg to 6.4 mmHg, approximately. This is expected, as for a small range of velocity the pressure drop across a particular obstruction will not vary significantly.

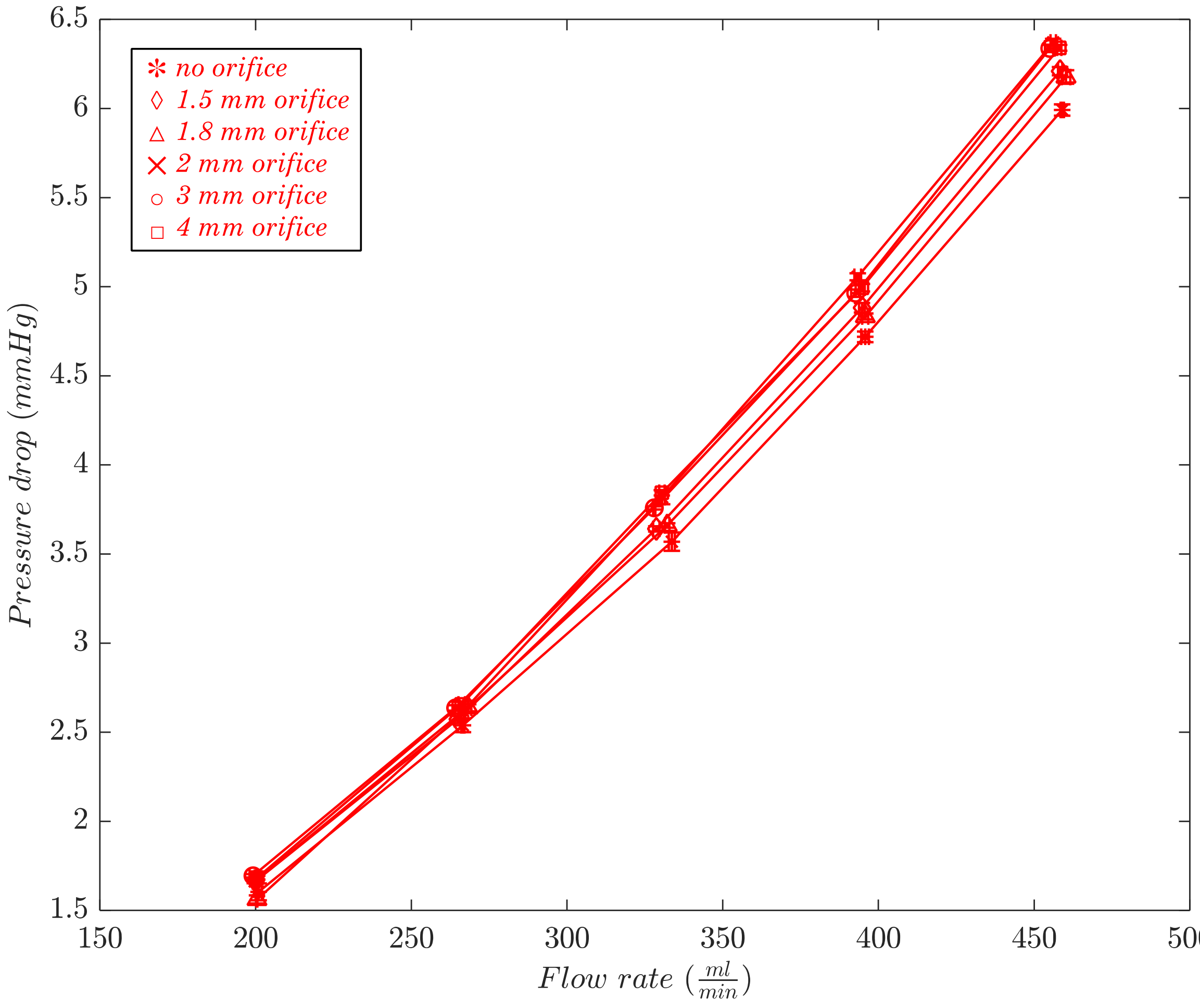

Figure A2: Pressure drop across the 75% stenosis for each of the downstream orifice configurations. * denotes with no orifice downstream, ◊ denotes an orifice of 1.5 mm diameter, △ denotes an orifice of 1.8 mm diameter, X denotes an orifice of 2 mm diameter, O denotes an orifice of 3 mm diameter, □ denotes an orifice of 4 mm diameter. Error bars are shown for both the pressure drop and the flow rate.

## Discrepancy analysis

Sensor accuracy is obtained from the datasheets of the sensors. For the pressure sensors, the accuracy is ±0.04% full scale. Because of the range of sizes used for the orifices downstream, there is a range of pressure values. Hence, two different ranges of pressure sensors were used to minimise sensor error. For test cases with 2, 3 and 4 mm orifices downstream, the pressure sensor range was 0 – 1 psi. For the test cases with 1.5 and 1.8 mm orifices downstream, the pressure sensor range was 0 – 5 psi. Thus, the accuracy of the pressure sensors is ±0.0004 and ±0.002 psi respectively. For the flow meter, the accuracy is 0.4% of the measured value plus 1.7 ml/min.

As mentioned in the experimental plan section, to minimise random errors, tests were repeated five times and justifiable outlier tests were discarded. Both the flow rate and pressure values were averaged and the standard deviation of the recorded values is presented as error bars in the figures in the results section.

The standard deviation in calculated values can be estimated using propagation of uncertainty formulae. Table A1 shows these equations (Skoog et al. 2007) and the calculated average uncertainty.

Table A1: Standard deviation equations for calculated values

| Equation | Standard deviation formula | Average uncertainty |
|---|---|---|
| $FFR = \frac{P_d}{P_a}$ | $\lvert FFR \rvert \sqrt{\left(\frac{\sigma_{Pd}}{P_d}\right)^2 + \left(\frac{\sigma_{Pa}}{P_a}\right)^2}$ | 0.0109 |
| $hSR = \frac{(P_a - P_d)_{mean}}{V_{mean}}$ | $\lvert hSR \rvert \sqrt{\frac{\sigma_{Pd}{}^2 + \sigma_{Pa}{}^2}{(P_d + P_a)^2} + \left(\frac{\sigma_{Vmean}}{V_{mean}}\right)^2}$ | 0.0043 |
| $hMR = \frac{P_d}{V_{mean}}$ | $\lvert hMR \rvert \sqrt{\left(\frac{\sigma_{Pd}}{P_d}\right)^2 + \left(\frac{\sigma_{Vmean}}{V_{mean}}\right)^2}$ | 0.0225 |
| $hMR_{pa} = \frac{P_a}{V_{mean}}$ | $\lvert hMR_{pa} \rvert \sqrt{\left(\frac{\sigma_{Pa}}{P_a}\right)^2 + \left(\frac{\sigma_{Vmean}}{V_{mean}}\right)^2}$ | 0.0266 |